\documentclass[12pt]{article}%
\usepackage{cite}
\usepackage{fix-cm}
\usepackage{color}
\usepackage[usenames,dvipsnames]{xcolor}
\usepackage{heck}
\usepackage{amsmath}
\usepackage{amsfonts}
\usepackage{amssymb}
\usepackage{graphicx}
\usepackage{multicol}
\usepackage{hyperref}
\usepackage{setspace}
\usepackage[all]{xy}
\usepackage{verbatim}
\usepackage{color}
\usepackage{epsfig}
\usepackage{mathtools}
\usepackage[margin=1in]{geometry}%
\providecommand{\U}[1]{\protect\rule{.1in}{.1in}}
\numberwithin{equation}{section}

\def\be{\begin{equation}}
\def\ee{\end{equation}}
\def\bea{\begin{eqnarray}}
\def\eea{\end{eqnarray}}

\definecolor{gray}{rgb}{0.5,0.5,0.5}
\begin{document}

\date{May 2013}

\title{Statistical Inference and String Theory}

\institution{HarvardU}{\centerline{Jefferson Physical Laboratory, Harvard University, Cambridge, MA 02138, USA}}

\authors{Jonathan J. Heckman\footnote{e-mail: {\tt jheckman@physics.harvard.edu}}}

\abstract{In this note we expose some surprising connections
between string theory and statistical inference.
We consider a large collective of agents sweeping out a family of
nearby statistical models for an $M$-dimensional
manifold of statistical fitting parameters. When the agents making nearby inferences
align along a $d$-dimensional
grid, we find that the pooled probability that the collective reaches a correct
inference is the partition function of a non-linear sigma model in $d$ dimensions.
Stability under perturbations to the original inference scheme requires the agents of the collective to distribute
along two dimensions. Conformal invariance
of the sigma model corresponds to the condition of a stable inference scheme,
directly leading to the Einstein field equations for classical gravity. By summing over all
possible arrangements of the agents in the collective, we reach a string theory. We
also use this perspective to quantify how much an observer can hope to
learn about the internal geometry of a superstring compactification. Finally, we present some
brief speculative remarks on applications to the AdS/CFT correspondence and Lorentzian signature
spacetimes.}

\enlargethispage{\baselineskip}

\setcounter{tocdepth}{2}

\renewcommand\Large{\fontsize{15}{17}\selectfont}

\maketitle

\tableofcontents

\newpage

\section{Introduction}

Sifting through competing interpretations of data lies at the core of
quantitative approaches to statistical inference. A successful statistical
model must strike a balance between the competing demands of accuracy and
simplicity. A related consideration is the ability to adapt an inference
scheme to new information.

In this note we show that this and related questions in statistical inference
are amenable to study using well-known results from string theory and quantum
field theory. Conversely, we use statistical inference to gain a different
perspective on string theory. Though we couch our results in broader terms,
one can also view this note as an attempt to define an approximate notion of a
local observable in quantum gravity.\footnote{In quantum gravity, it is hard
to define a local off-shell observable because selecting a point of the spacetime breaks
diffeomorphism invariance. For spacetimes which asymptote to either Minkowski
or Anti-de Sitter space, more precision is available because observables can
be formulated in terms of boundary data.} In the specific context of string theory, there is
a related issue as to what are the underlying principles
which require the presence of strings in some regime of validity. In a rather
unexpected way, classical gravity and perturbative strings will indeed emerge from our considerations.

In more formal terms, we frame the question of statistical inference as the
attempt of an agent to develop a statistical model after observing independent
events $E=\left\{  e_{1},...,e_{N}\right\}  $ which have been drawn from the
true\ probability distribution. The task of the agent is to produce an
accurate statistical model $A$ depending on $M$ continuous statistical
parameters $y^{(1)},...,y^{(M)}$. Given two competing models $A$ and $B$, we
can then compare the Bayesian posterior probabilities $\Pr(A|E)$ and
$\Pr(B|E)$, and select the model with the higher value.

In \cite{BalasubramanianGeo, Balasubramanian:1996bn}, the value of $\Pr(A|E)$
was interpreted as the partition function of a statistical mechanics problem.
The parameters of the statistical model specify the configuration space of a
thermodynamic ensemble, with $N$ playing the role of an inverse temperature,
and the proximity of the guess from the true distribution playing the role of
an energy. The competition between decreasing the energy and increasing the
entropy of the ensemble translates to the competing interests of achieving a
better fit to the data, but with as simple a model as possible. This
thermodynamic interpretation is rather striking, and suggests a number of generalizations.

Our aim in this note is to generalize this analysis to a collective of agents,
who may decide to use different statistical models to fit the data. Since the
collective gets to explore many nearby models, it can infer a broader class of
inference strategies, and in particular, may arrive at a different inference
than any individual agent. An additional reason to consider a collective is
that its inferences may be more stable against perturbations.

We define a collective as a large number of agents $K\equiv K_{\text{agent}}$
with statistical models $A_{1},...,A_{K}$, each of which depend on the same
$M$ parameters $y^{(1)},...,y^{(M)}$, which can be correlated across different agents.
Each member of the collective samples the true distribution, and receives
a set of $N$ events $E_{(i)}=\{e_{1}^{(i)},...,e_{N}^{(i)}\}$ for $i=1,...,K$.
The pooled posterior probability of the collective is defined as:
\begin{equation}
Z(A_{\text{coll}}|E_{\text{coll}})\equiv\Pr(A_{1}|E_{1})...\Pr(A_{K}|E_{K}).
\end{equation}
The success of the collective versus the individual agent is then obtained by
comparing the geometric mean $[Z(A_{\text{coll}}|E_{\text{coll}})]^{1/K}$ with
$\Pr(A|E)$.

Depending on the nature of the true distribution, the collective could decide
on various inference strategies. For example, given two collectives
$A_{\text{coll}}$ and $B_{\text{coll}}$, collective $A_{\text{coll}}$ could
consist of agents trying to fit to a Gaussian, as well as other distributions
which are small deformations of a Gaussian profile. Alternatively, collective
$B_{\text{coll}}$ could try fitting to a Lorentzian, and nearby distributions.
The aim is to pick the collective with the higher pooled posterior probability.

In a similar vein to \cite{BalasubramanianGeo, Balasubramanian:1996bn}, our
aim will be to interpret $Z(A_{\text{coll}}|E_{\text{coll}})$ in terms of a
statistical mechanical model. When the guesses of nearby agents are arranged
along a $d$-dimensional grid, the geometry of agents builds up an
approximation to a smooth manifold $\Sigma_{\text{agent}}$ with coordinates
$\sigma^{1},...,\sigma^{d}$. Each agent makes a guess $y^{(I)}(\sigma^{a})$
for $I=1,...,M$ and $a=1,...,d$. Varying over the choice of $\sigma^{a}$ on
the grid then yields a family of nearby guesses. In the limit $K\gg1$ where
the agents fill in a dense mesh, the $y$'s define a map from the space of
agents to the space of parameters:%
\begin{equation}
y:\Sigma_{\text{agent}}\rightarrow Y_{\text{target}},
\end{equation}
so for each point of $\Sigma_{\text{agent}}$, we get an agent with a
corresponding statistical model. The defining property of the
$d$-dimensional collective is that nearby guesses are correlated:
\begin{equation}
\delta y^{I} \delta y^{J} \rightarrow h^{a b} \partial_{a} y^{I} \partial_{b} y^{J}
\end{equation}
where $h^{a b}$ is a possibly position dependent positive definite $d \times d$ matrix which
defines a notion of proximity between agents on the grid.

We find that in the large $N$ limit, the probability $Z(A_{\text{coll}%
}|E_{\text{coll}})$ is a path integral:%
\begin{equation}
Z(A_{\text{coll}}|E_{\text{coll}})\propto\int\left[  \mathcal{D}y\right]
\text{ }\sqrt{\det G}\text{ }e^{-S_{\text{tot}}},
\end{equation}
up to a normalization constant which we shall mostly neglect. Here, $\left[
\mathcal{D}y\right]  $ $\sqrt{\det G}$ is a measure factor for the path
integral, i.e., a measure for the space of all maps $y:\Sigma_{\text{agent}%
}\rightarrow Y_{\text{target}}$. The exponent $S_{\text{tot}}$ is the action
of a non-linear sigma model in $d$ Euclidean dimensions:
\begin{equation}
S_{\text{tot}}=\underset{\Sigma_{\text{agent}}}{\int}d^{d}\sigma\text{ }%
\sqrt{\det h}\text{ }\left(  \frac{N}{2}G_{IJ}\text{ }h^{ab}\frac{\partial
y^{I}}{\partial\sigma^{a}}\frac{\partial y^{J}}{\partial\sigma^{b}}%
+V(y,\sigma)\right)
\end{equation}
and $G_{IJ}$ is an information metric\ on the space of statistical parameters
$Y_{\text{target}}$.\footnote{The study of differential geometry on
statistical parameter spaces is known as information geometry. See
\cite{AmariNag} for an excellent review.} Here and throughout, repeated
subscripts and superscripts are to be summed.

The information metric is defined by picking a family of probability
densities $p_{A}(x,\{y\})$ and varying the parameters $y$:
\begin{equation}
G_{IJ}=\underset{X}{\int}d\mu(x)\text{ }p_{A}\frac{\partial\log p_{A}%
}{\partial y^{I}}\frac{\partial\log p_{A}}{\partial y^{J}}.\label{Fisher}%
\end{equation}
This measures the infinitesimal proximity of $p_{A}$ to the true distribution.

The kinetic term of the sigma model is simply the pullback of the information
metric on $Y_{\text{target}}$ to $\Sigma_{\text{agent}}$. Summing over all the
agents yields a notion of proximity of the true distribution to the guessing
strategy adopted by the entire collective.

Adding a potential energy function to the non-linear sigma model corresponds
to a choice of statistical prior. This allows for local interactions between
nearby neighbors on the grid. Such deformations can be either Lorentz
invariant or possibly Lorentz breaking, signifying a preferred weighting of
specific agents in the collective.

\begin{figure}
[ptb]
\begin{center}
\includegraphics[
height=1.74in,
width=2.936in
]%
{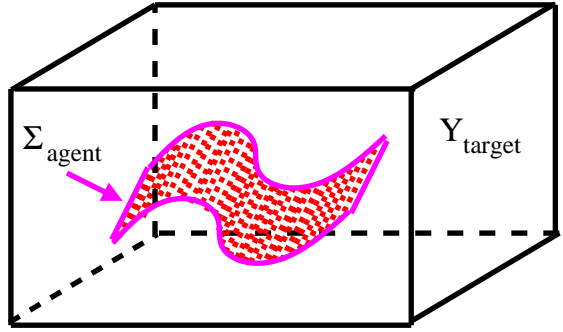}%
\caption{Depiction of a collective of agents arranged on a $d$-dimensional
grid of $\Sigma_{\text{agent}}$ making nearby guesses of statistical fitting
parameters on an $M$-dimensional geometry $Y_{\text{target}}$.}%
\label{stattarget}%
\end{center}
\end{figure}

The $d$-dimensional field theory specifies a collection of agents making
nearby guesses on a spatial $d$-dimensional grid. If we distribute the agents
on a grid with a Lorentzian signature metric, we can alternatively view this
as a collective of agents on a ($d-1$)-dimensional grid who update their
inferences in discrete time steps. See figure \ref{stattarget} for a
depiction.%

The field theory formulation provides a tool for addressing how the collective
would respond to a perturbation to the original inference, and moreover, how
estimates on the parameters are correlated over the lattice of agents. If we
add a linear source term $\mathcal{J}_{I}(\sigma)y^{I}(\sigma)$, we can
specify initial conditions for some of the agents. Then, the one-point
function will tell us about how the original inference strategy settles
towards a preferred value of parameters. The higher point functions tell us
about correlations between the guesses of nearby agents.

We can also ask what happens if there is a perturbation to the family of
statistical models used by the collective. One might view this as the
collective \textquotedblleft changing its mind\textquotedblright, or
equivalently, reacting to a change in the true distribution. Ideally, such
perturbations should not completely destabilize the original inference, though
the collective should be capable of altering its original inference. For
simplicity, we mainly focus on the case of uniformly weighted agents in flat space.

Surprisingly, the answer is very sensitive to the geometry of the agents. For
$d<2$, a small change in the probability distribution generically destabilizes
the original inference, that is, the collective does not reach the same
inference. For $d>2$, these perturbations wash out, and do not change the
original inference. When $d=2$, however, the collective can adjust its
inference procedure.

In other words, stable statistical inference selects out two dimensions for
the grid of agents. The limiting case where the overall dependence on the
number of agents drops out translates to the condition of conformal
invariance in the sigma model. As is well known the condition of conformal
invariance leads directly to the Einstein
field equations for classical gravity. Quantum fluctuations around the
background metric arise from fluctuations in the inferred probability
distribution. As far as we are aware, this is the first derivation of
classical gravity from the condition of stable statistical inference.

The classical description breaks down at small distance in the fitting
parameter space. This occurs at a statistical resolution length (in units
where $2\pi\alpha^{\prime}=1/N$):%
\begin{equation}
\ell_{\text{min}}\sim1/\sqrt{N}%
\end{equation}
with $N$ the number of sampled events. In physics terms, this sets a cutoff
for the target space theory, which we interpret as the string scale associated
with the sigma model.

To reach a full string theory, we can make the stronger demand that we sum
over all possible arrangements of the agents by coupling the sigma model to
two-dimensional gravity. Quantum stability can be ensured by passing to the
superstring. Each agent can then be viewed as the endpoint of an open string,
that is, a D-instanton, and a collective of D-instantons can build up a string.
From this perspective, the condition of stable inference gives an explanation for
why a quantum theory of gravity must contain stringlike excitations.

Turning the discussion around, we also apply methods from statistical
inference to the study of superstring theory. We ask how well an observer
could hope to resolve the internal geometry of a compactified spacetime
$\mathbb{R}^{3,1}\times Y_{\text{internal}}$. A guess by an observer defines a
probability distribution which depends on the location of the observer $y\in
Y_{\text{internal}}$, as well as other fitting parameters which characterize
the local profile of the metric. This generates a family of probability
distributions, and a corresponding information metric. Calabi-Yau
compactifications can be understood in information theoretic terms as
generating successively better numerics for a balanced metric (see e.g.
\cite{Tian, Zelditch, DonaldsonApprox, Douglas:2006rr}). Comparing the true
probability distribution with the original guess, we can then quantify the
amount of information an agent would gain by adjusting their initial incorrect
guess of the internal geometry. We also entertain some more speculative
material on potential applications to the AdS/CFT correspondence, as well as
Lorentzian signature statistical manifolds.

The rest of this note is organized to reflect the fact that some pieces may be
of interest to different readers. In section \ref{sec:REVIEW} we briefly
review some background on statistical inference and information geometry, and
in section \ref{sec:FIELD} we present a sigma model for statistical inference
and study some of its properties using results which --though well-known to
string theorists-- may seem surprising. After this, we focus our attention on
applications to the physical superstring. We study the information content of
a superstring compactification in section \ref{sec:COMPACT}, and briefly
entertain some more speculative applications in section \ref{sec:SPEC}. We
conclude in section \ref{sec:CONC}.

\section{Statistical Inference and Information Geometry \label{sec:REVIEW}}

In this section we review the main concepts from statistical inference and
information geometry which we shall use. Excellent resources are the reviews
\cite{CoverThomas, AmariNag}.

Given a probability space $X$ with measure $\mu(x)$ we would like to define
some notion of two probability distributions being nearby. A common choice is
the Kullback-Leibler divergence between the true density $p_{T}(x)$ and a
guess $p_{A}(x)$:%
\begin{equation}
D_{KL}(p_{T}||p_{A})\equiv\underset{X}{\int}d\mu(x)\text{ }p_{T}(x)\log
\frac{p_{T}(x)}{p_{A}(x)},
\end{equation}
which is also known as the relative entropy. Though it is not a true metric
(as it is not symmetric), the Kullback-Leibler divergence is non-negative, and
has a simple statistical interpretation as quantifying the amount of
information (in the sense of Shannon \cite{Shannon:1948zz}) an agent would
gain by learning the true distribution. One can also introduce the continuum
generalization of the Shannon entropy known as the differential entropy, by
integrating $-p\log p$. This must be treated with more care, because this
\textquotedblleft entropy\textquotedblright\ can be negative.

Of course, the Kullback-Leibler divergence is but one way to compare the
proximity of distributions. In fact, in the small distance limit,
other notions of proximity produce the same notion of distance between
distributions. To see this, consider a parametric family of
probability densities $p_{A}(x,\left\{  y\right\}  )$ depending on
coordinates $\left\{  y^{I}\right\}  $ of some manifold $Y_{\text{target}}$:%
\begin{equation}
p_{T}=p_{A}+\delta\equiv p_{A}+\delta y^{I}\frac{\partial p_{A}}{\partial
y^{I}}\label{deltaapprox}%
\end{equation}
which are infinitesimally close to the true distribution. Expanding to second
order in $D_{KL}$ then yields the Fisher information metric:%
\begin{equation}
D_{KL}(p_{A}+\delta||p_{A})\simeq G_{IJ}\text{ }\delta y^{I}\delta
y^{J}+O(\delta y^{3})\label{inf}%
\end{equation}
with $G_{IJ}$ as in equation (\ref{Fisher}).

A central result in information geometry is Chentsov's theorem \cite{Chentsov}.
This theorem guarantees that at least in situations where we can sample the
true distribution sufficiently often, the infinitesimal form of the distance
will eventually converge to the Fisher information metric, that is, equation
(\ref{inf}). In this sense, the infinitesimal form of statistical inference is
relatively insensitive to which finite distance measure between distributions
we choose to adopt.

Now, having introduced a metric $G_{IJ}$ for the Riemannian manifold
$Y_{\text{target}}$, the first inclination of a physicist might be to study
geodesic flows with respect to the Levi-Cevita connection $\nabla_{LC}$. This
is certainly possible to do, but it turns out to not be the only natural
choice\ in the case of statistical inference. We can also introduce
connections with torsion $\nabla_{B}$ which are specifically adapted to a
fixed choice of affine coordinate system. We can then define another
connection $\nabla_{B}^{\ast}$ via the implicit relation $\nabla_{B}%
+\nabla_{B}^{\ast}=2\nabla_{LC}$. The use of connections with torsion is quite
prevalent in the information geometry literature. This is because one is often
concerned with a privileged class of probability distributions, and therefore
must adapt an affine coordinate system for this specific distribution.
Parallel transport with respect to the Levi-Cevita connection will generically
disrupt this coordinate system, so it is common (see e.g. \cite{AmariNag}) to
label a statistical manifold by the triple $(G_{IJ},\nabla_{B},\nabla
_{B}^{\ast})$.

To illustrate some of the above notions, consider the case of a normal
probability distribution for outcomes $x^{(I)}\in%
\mathbb{R}
^{M-1}$. The fitting parameters are $M-1$ positions and an overall width:%
\begin{equation}
p(x;\left\{  y^{(1)},...,y^{(M-1)},\alpha\right\}  )=\left(  \pi\alpha
^{2}\right)  ^{-(M-1)/2}\exp\left[  -\underset{I=1}{\overset{M-1}{\sum}}%
\frac{(x^{(I)}-y^{(I)})^{2}}{\alpha^{2}}\right]  \label{normal}%
\end{equation}
The information metric is:
\begin{equation}
ds^{2}=2\times\frac{dy_{(I)}dy^{(I)}+d\alpha^{2}}{\alpha^{2}} \label{metric}%
\end{equation}
We find it striking that this is also the metric for Euclidean
Anti-de Sitter space in $M$ dimensions.

\subsection{Statistical Mechanics of Statistical Inference}

We now review the statistical mechanical interpretation of statistical
inference \cite{BalasubramanianGeo, Balasubramanian:1996bn}. We start with a
probability distribution $T$ with density $p_{T}(x)$ which has generated $N$
independent events $E=\left\{  e_{1},...,e_{N}\right\}  $. The aim is to find
a statistical model which comes closest to approximating this distribution.
Given two models $A$ and $B$ we can compute $\Pr(A|E)$ and $\Pr(B|E)$. The aim
is to pick the model with the higher conditional probability.

Suppose our agent has made a choice for a parametric family of models $A$ with $M$ continuous fitting parameters
$y^{(1)},...,y^{(M)}$. Using Bayes' rule, the Bayesian posterior probability
is:
\begin{equation}
\Pr(A|E)=\frac{\Pr(A)}{\Pr(E)}\frac{\int d^{M}y\text{ }w(y)\text{ }\Pr
(E| \{y \})}{\int d^{M}y\text{ }w(y)},
\end{equation}
where $w(y)$ is some choice of weighting scheme for the parameters $y^{(I)}$.
Here, $\Pr(A)$ is the prior probability for the family of models $A$, and $\Pr(E)$ is the
probability of the event set $E$.

Now, the key step is that in the limit where the agent has sampled a large
$N\gg1$ number of independent events, the law of large numbers implies that
$\Pr(E| \{y \})$ approaches (in the almost sure sense) \cite{BalasubramanianGeo,
Balasubramanian:1996bn}:
\begin{equation}
\Pr(E| \{y \})\simeq\exp(-N(\mathcal{E}_{A}+\mathcal{E}_{0}))
\end{equation}
where $\mathcal{E}_{A}$ is the Kullback-Leibler divergence and $\mathcal{E}%
_{0}$ is the differential entropy:%
\begin{equation}
\mathcal{E}_{A}=\underset{X}{\int}d\mu(x)\text{ }p_{T}(x)\log\frac{p_{T}%
(x)}{p_{A}(x , \{y \})}\text{ \ \ and \ \ }\mathcal{E}_{0}=-\underset{X}{\int}%
d\mu(x)\text{ }p_{T}(x)\log p_{T}(x)\text{.}%
\end{equation}
In other words, in the large $N$ limit $\Pr(A|E)$ is a partition function:
\begin{equation}
\Pr(A|E)=\frac{\Pr(A)}{\Pr(E)}\frac{\int d^{M}y\text{ }w(y)\text{
}e^{-N\left(  \mathcal{E}_{A}+\mathcal{E}_{0}\right)  }}{\int d^{M}y\text{
}w(y)}. \label{priorguessing}%
\end{equation}

This provides a statistical mechanical interpretation for statistical
inference \cite{BalasubramanianGeo, Balasubramanian:1996bn}. The number of
sampled events $N$ corresponds to the inverse temperature, and the
Kullback-Leibler divergence $\mathcal{E}_{A}$ is an energy. Up to a
$y$-independent shift, the quantity $\mathcal{E}_{0}$ can be repackaged as
indicating the interaction between the guess and \textquotedblleft statistical
defects\textquotedblright. Moreover, just as in thermodynamics, there is a
competition between minimizing the energy, and maximizing the entropy. This
corresponds to balancing the interests of minimizing the distance between
$p_{A}$ and $p_{T}$, and on the other hand, the possibility of probing more
nearby configurations. In the \textquotedblleft quenched
approximation\textquotedblright\ to the partition function, we can basically
drop the background value of $\mathcal{E}_{0}$. Indeed, we shall mostly
neglect this contribution in our considerations.

If we assume that the probability density $p_{A}(x,\left\{  y\right\}  )$ is
nearby the true density in the sense of equation (\ref{deltaapprox}), we can
further approximate the energy as:
\begin{equation}
\mathcal{E}_{A}\simeq G_{IJ}\text{ }\delta y^{I}\delta y^{J}+O(\delta y^{3}).
\label{EGapprox}%
\end{equation}

Finally, in \cite{BalasubramanianGeo, Balasubramanian:1996bn} additional care
is paid to the choice of a weighting function $w(y)$, where it is argued that
the natural choice for a uniform sampling of models is the Jeffrey's prior,
corresponding to the choice $w(y)=\sqrt{\det G}$. The integration measure
is then parametrization invariant. Of course, depending on the prior
information, one might make different choices.

\section{Field Theory for Statistical Inference \label{sec:FIELD}}

In this section we turn to the core of our analysis, where we generalize
\cite{BalasubramanianGeo, Balasubramanian:1996bn} to a collective of agents.
Now, whereas a single agent must settle for one inference scheme, a collective
can sweep out a bigger set of guesses. What we show is that the pooled
posterior probability of the collective fitting the data is the partition
function of a non-linear sigma model. We include some additional heuristic
background on such field theories since some aspects may be unfamiliar to the reader.

We define a collective $A_{\text{coll}}$ of $K$ agents as a set of nearby
statistical models $A_{1},...,A_{K}$. Each agent draws $N$ events from the
true distribution $E_{i}=\left\{  e_{1}^{(i)},...,e_{N}^{(i)}\right\}  $ for
$i=1,...,K$. The models are assumed to be nearby in the sense that they each
depend on $M$ parameters $y^{(1)},...,y^{(M)}$ for some guess distribution. In
other words, we get $K$ sets of fitting parameters $y_{1},...,y_{K}$, where
each $y_{i}$ itself specifies $M$ fitting parameters. Restoring all indices,
we label each such parameter as $y_{j}^{(I)}$ for $j=1,...,K$ and $I=1,...,M$.
To be part of the collective, we assume that the fitting parameters of agents are correlated. The
specific type of correlation between agents is taken as part of the defining data of the collective.

For each agent, we have the Bayesian posterior probability:
\begin{equation}
\Pr(A_{i}|E_{i})=\frac{\Pr(A_{i})}{\Pr(E_{i})}\frac{\int d^{M}y_{i}\text{
}w_{i}(y_{i})\text{ }e^{-N\left(  \mathcal{E}_{A}^{(i)}+\mathcal{E}_{0}%
^{(i)}\right)  }}{\int d^{M}y_{i}\text{ }w_{i}(y_{i})}%
\end{equation}
where $\mathcal{E}_{A}^{(i)}$ is the energy for each agent, and $\mathcal{E}%
_{0}^{(i)}$ is the differential entropy of the true distribution. In
principle, the prior probability $\Pr(A_{i})$ of each agent's guess, as well
as their weighting schemes $w_{i}(y_{i})$ could vary from agent to agent.

Our interest is not in the success of any individual agent, but instead the
full collective. What we want to study is the pooled posterior probability of
the collective as a whole to reach some inference. Along these lines, we
introduce the pooled posterior probability of the collective:%
\begin{equation}
Z(A_{\text{coll}}|E_{\text{coll}})\equiv\underset{i=1}{\overset{K}{%
{\displaystyle\prod}
}}\Pr(A_{i}|E_{i}). \label{Zcollective}%
\end{equation}
When the context is clear, we shall denote this by $Z_{\text{coll}}$. This
definition reflects the assumption that although the agents of the collective
are required to make nearby guesses, they are otherwise independent. Given two
collectives $A_{\text{coll}}$ and $B_{\text{coll}}$ each composed of $K$
agents, our aim will be to select the collective with the higher pooled
posterior probability. We can also compare the success of a collective to that
of a single agent by computing $[Z(A_{\text{coll}}|E_{\text{coll}})]^{1/K}$.

\subsection{Lattice Approximation}

To give the problem some more structure, we shall now assume that the agents
are aligned along a $d$-dimensional grid, which is assumed to be a lattice
approximation to some $d$-dimensional manifold $\Sigma_{\text{agent}}$. For
each choice of point in the lattice approximation to $\Sigma_{\text{agent}}$,
we can pick a value for $y^{I}(\sigma^{a})$. In other words, we build up an
approximation to the map:%
\begin{equation}
y:\Sigma_{\text{agent}}\rightarrow Y_{\text{target}}\text{.}%
\end{equation}
In this case, the collective is \textit{defined} by the condition that
nearby fluctuations in the agent space are correlated as:
\begin{equation}
\delta y^{I} \delta y^{J} \rightarrow h^{a b} \frac{\partial y^{I}}{\partial \sigma^{a}} \frac{\partial y^{J}}{\partial \sigma^{b}}
\end{equation}
with summation on the indices $a,b = 1,...,d$ implicit.
In this case, performing the integral $\underset{i=1}{\overset{K}{%
{\displaystyle\prod}
}}\int d^{M}y_{i}$ then sweeps out a range of possible maps.

Let us illustrate the lattice approximation for $\Sigma_{\text{agent}%
}=\mathbb{R}^{d}$. We introduce a collection of $d$ unit normalized vectors
$\overrightarrow{e}_{1}=(1,0,...,0)$,..., $\overrightarrow{e}_{d}%
=(0,...,0,1)$. We span a grid with the linear combinations $\ell\underset
{a=1}{\overset{d}{\sum}}m_{a}\overrightarrow{e}_{a}$ where $\ell$ is a small
parameter and $m_{a}$ are integers. A lattice derivative in the direction
$\overrightarrow{e}_{a}$ corresponds to:%
\begin{equation}
\delta_{a}y^{J}=\frac{y^{J}(\overrightarrow{\sigma}+\ell\overrightarrow{e}%
_{a})-y^{J}(\overrightarrow{\sigma})}{2\ell}+\frac{y^{J}(\overrightarrow
{\sigma})-y^{J}(\overrightarrow{\sigma}-\ell\overrightarrow{e}_{a})}{2\ell
}\label{latticederivative}%
\end{equation}
for $a=1,...,d$ and $J=1,...,M$. Discretized integration corresponds to the
substitution:%
\begin{equation}
\ell^{d}\text{ }\underset{i=1}{\overset{K}{\sum}}\rightarrow\int d^{d}%
\sigma\text{ }\sqrt{\det h}%
\end{equation}
where we have introduced a notion of distance between neighboring agents via
the metric $ds_{\text{agent}}^{2}=h_{ab}d\sigma^{a}d\sigma^{b}$ on the agent
space. To focus on the essential points, we shall typically take $h_{ab}$ to
be a constant matrix. Much of what we say generalizes to a broader class of
agent space metrics.

In practice, the space $\Sigma_{\text{agent}}$ will typically have finite
volume. The lattice approximation therefore amounts to fixing some small unit
cell of size $\ell^{d}$, and arranging these cells to produce the total volume
of $\Sigma_{\text{agent}}$:%
\begin{equation}
\text{Vol}(\Sigma_{\text{agent}})=K\times\ell^{d}.
\end{equation}
The continuum limit corresponds to sending $\ell\rightarrow0$ and
$K\rightarrow\infty$ with Vol$(\Sigma_{\text{agent}})$ held fixed.

We would now like to argue that inference of the collective defines a
$d$-dimensional quantum field theory of a very particular type. Since we are
assuming the family of guesses are close to the true distribution, we can
expand the energy at a point $\left\{  \sigma^{a}\right\}  $ as in equation
(\ref{EGapprox}). This defines a pullback of the metric on $Y_{\text{target}}$
to a metric on $\Sigma_{\text{agent}}$:%
\begin{equation}
\mathcal{E}_{A}(\sigma)\simeq G_{IJ}\text{ }\delta y^{I}\delta y^{J}\simeq
G_{IJ}\text{ }h^{ab}\frac{\partial y^{I}}{\partial\sigma^{a}}\frac{\partial
y^{J}}{\partial\sigma^{b}}%
\end{equation}
where $h^{ab}$ denotes the matrix inverse of $h_{ab}$. In the limit of a large
number of agents, the lattice approximation tends to:%
\begin{equation}
\underset{i=1}{\overset{K}{%
{\displaystyle\prod}
}}\int d^{M}y_{i}\text{ }e^{-N\mathcal{E}_{A}^{(i)}}\rightarrow\int\left[
\mathcal{D}y\right]  e^{-S_{\text{kin}}}%
\end{equation}
up to a common overall constant normalization factor. Here, the measure factor
$\left[  \mathcal{D}y\right]  $ is a heuristic instruction to integrate over
all possible maps $\Sigma_{\text{agent}}\rightarrow Y_{\text{target}}$. The
contribution $S_{\text{kin}}$ is the kinetic term for a non-linear sigma
model:%
\begin{equation}
S_{\text{kin}}=\frac{1}{\ell^{d}}\int d^{d}\sigma\text{ }\sqrt{\det h}\text{
}\frac{N}{2}\text{ }G_{IJ}\text{ }h^{ab}\frac{\partial y^{I}}{\partial
\sigma^{a}}\frac{\partial y^{J}}{\partial\sigma^{b}}.
\end{equation}

The kinetic energy defines a sum over the infinitesimal Kullback-Leibler
divergences for all of the agents, measuring the proximity of the collective
to the true distribution. This provides a first indication of how a collective
inference might differ from an individual inference:\ Whereas individual
agents might produce sharp jumps in making different inferences, the
collective might instead smooth these out to only gradual changes across the
entire agent space. In statistical terms, the collective gains more by
exploring many nearby options than by trying to minimize the Kullback-Leibler
divergence of a single agent.

The other terms of equation (\ref{Zcollective}) also define natural objects in
this quantum field theory. For example, the agent dependent weighting factor
$w_{i}(y_{i})$ lifts to a possibly position dependent potential energy:%
\begin{equation}
\underset{i=1}{\overset{K}{%
{\displaystyle\prod}
}}w_{i}(y_{i})\rightarrow e^{-S_{\text{pot}}}%
\end{equation}
where:%
\begin{equation}
S_{\text{pot}}=-\frac{1}{\ell^{d}}\underset{\Sigma_{\text{agent}}}{\int}%
d^{d}\sigma\text{ }\sqrt{\det h}\text{ \ }\log(w(y,\sigma)),\label{Spot}%
\end{equation}
and similar considerations apply for the position dependent potential energy
on agent space:%
\begin{equation}
\underset{i=1}{\overset{K}{%
{\displaystyle\prod}
}}\frac{\Pr(A_{i})}{\Pr(E_{i})}e^{-N\mathcal{E}_{0}^{(i)}}\rightarrow
e^{-U_{\text{pos}}}.
\end{equation}
Again, this makes intuitive sense; a choice of weighting scheme dictates where
the statistical inference procedure may be attracted. A position dependent
potential term means not all agents are weighted equally. In most uniform
weighting schemes one is essentially demanding Lorentz invariance of the
continuum theory.

Assembling all of the pieces and integrating over the class of all possible
maps, i.e. choices of $y\left(  \sigma\right)  $, we have arrived at the path
integral formulation of a non-linear sigma model. Statistical inference of the
collective determines a quantum field theory!

It is convenient to adhere to standard practice in field theory by rescaling
the fields to eliminate the explicit factors of $\ell$ appearing in our
continuum theory integrals. Doing so, we arrive at the final form of the
partition function for a non-linear sigma model:%
\begin{equation}\label{Zcoll}
Z_{\text{coll}}\propto\int[\mathcal{D}y]\sqrt{\det G}\text{ }e^{-S_{\text{tot}%
}}.
\end{equation}
where:%
\begin{equation}
S_{\text{tot}}=\int d^{d}\sigma\text{ }\sqrt{\det h}\text{ }\left(  \frac
{N}{2}G_{IJ}\text{ }h^{ab}\frac{\partial y^{I}}{\partial\sigma^{a}}%
\frac{\partial y^{J}}{\partial\sigma^{b}}+V(y,\sigma)\right)
\end{equation}
in the obvious notation. This is the action in $d$ Euclidean dimensions for a
non-linear sigma model which has been deformed by a field and position
dependent potential energy. Here, we have adjusted the measure of the path
integral by a factor of $\sqrt{\det G}$ so that $V=0$ refers to the Jeffrey's prior.

\subsubsection*{A More Formal Derivation}

At various stages in our derivation, we took some heuristic liberties. In this
subsection we give a more formal derivation. We can view the $K$ agents as a
single \textquotedblleft meta-agent\textquotedblright\ which has sampled the
probability space $X_{(1)}\times...X_{(K)}$. We label an outcome from this
probability space as the $K$-tuple $\mathcal{X}=\left\{  x_{(1)}%
,...,x_{(K)}\right\}  $. The entire set of fitting parameters is then
$\{\mathcal{Y}^{\mathcal{I}}\}=\left\{  y_{(1)}^{I},...,y_{(K)}^{I}\right\}  $
for $I=1,...,M$. The index $\mathcal{I}$ indicates both an agent $i=1,...,K$
as well as a fitting index $I=1,...,M$. In this formalism, the true
distribution sampled by the agents of the collective is simply the product:%
\begin{equation}
\mathcal{P}_{T}(\mathcal{X})=\underset{i=1}{\overset{K}{%
{\displaystyle\prod}
}}p_{T}(x_{(i)}),
\end{equation}
and the parametric family of models used by the collective can be summarized
as $\mathcal{P}_{A}(\mathcal{X};\{\mathcal{Y}\})$. The inference scheme of the
collective can be viewed as a single meta-agent drawing $N$ events from the
true distribution $\mathcal{P}_{T}(\mathcal{X})$. Each event is itself a
$K$-tuple of events drawn from the $p_{T}(x_{(i)})$. Denote this set of events
by $E_{\text{meta}}$, and let $A_{\text{meta}}$ denote the family of models
parameterized by the $\{\mathcal{Y}\}$. Then, we can re-write the pooled
posterior probability as:
\begin{equation}
Z\left(  A_{\text{coll}}|E_{\text{coll}}\right)  =\Pr(A_{\text{meta}%
}|E_{\text{meta}})=\frac{\Pr(A_{\text{meta}})}{\Pr(E_{\text{meta}})}\frac{\int
d\mathcal{Y}\text{ }\mathcal{W}(\mathcal{Y})\Pr(E_{\text{meta}}|\{\mathcal{Y}%
\})}{\int d\mathcal{Y}\text{ }\mathcal{W}(\mathcal{Y})}.
\end{equation}
In physics terms, the integration measure $d\mathcal{Y}=[\mathcal{D}y]$ is
simply that used in the path integral.

To define the notion of a grid, we will actually need to enlarge the fitting
parameter space to $d$ copies of the meta-parameters. So, we label these
fitting variables as $\mathcal{Y}_{a}^{\mathcal{I}}$, where $a=1,...,d$ runs
over the dimensions of the grid. To enforce the condition that there are no
additional degrees of freedom in the fit, we assume that $\mathcal{W}%
(\mathcal{Y})$ includes a delta function constraint which imposes the
condition $\mathcal{Y}^{\mathcal{I}}=\mathcal{Y}_{1}^{\mathcal{I}}%
=\mathcal{Y}_{2}^{\mathcal{I}}=...=\mathcal{Y}_{d}^{\mathcal{I}}$.

Now, using the law of large numbers to convert $\Pr(E_{\text{meta}%
}|\{\mathcal{Y}\})$ into a Boltzmann factor, we can express the conditional
probability $\Pr(E_{\text{meta}}|\{\mathcal{Y}\})$ as:%
\begin{equation}
\Pr(E_{\text{meta}}|\{\mathcal{Y}\})\simeq\exp\left(  -N\text{ }%
\mathcal{G}_{\mathcal{IJ}}^{ab}\delta\mathcal{Y}_{a}^{\mathcal{I}%
}\mathcal{\delta Y}_{b}^{\mathcal{J}}\right)  .
\end{equation}
where $\mathcal{G}_{\mathcal{IJ}}^{ab}$ is the Fisher information metric for
the meta-agent:%
\begin{equation}
\mathcal{G}_{\mathcal{IJ}}^{ab}=\int d\mu(\mathcal{X})\text{ }\mathcal{P}%
_{A}\frac{\partial\log\mathcal{P}_{A}}{\partial\mathcal{Y}_{a}^{\mathcal{I}}%
}\frac{\partial\log\mathcal{P}_{A}}{\partial\mathcal{Y}_{b}^{\mathcal{J}}%
}.\label{generalized}%
\end{equation}
The energy functional $\mathcal{G}_{\mathcal{IJ}}^{ab}\delta\mathcal{Y}%
_{a}^{\mathcal{I}}\mathcal{\delta Y}_{b}^{\mathcal{J}}$ can now be viewed as a
matrix product in three different spaces. First, there is the sum over $a,b$,
i.e. the different directions of the grid. Second, there is the meta index
$\mathcal{I}$ which splits up as a choice of agent, and a direction $I$ in the
fitting parameter space. The simplest possibility which is generated by
product distributions is that $\mathcal{G}_{\mathcal{IJ}}^{ab}$ simply factors
as:%
\begin{equation}
\mathcal{G}_{\mathcal{IJ}}^{ab}=G_{IJ}h^{ab}\delta^{2}(\sigma-\sigma^{\prime
}),
\end{equation}
that is, we sum over $I,J=1,...,M$, $a,b=1,...,d$, and integrate over
$\sigma,\sigma^{\prime}$ in the agent space. The variation:
\begin{equation}
\delta\mathcal{Y}_{a}^{\mathcal{I}}\rightarrow\partial_{a}\mathcal{Y}%
^{\mathcal{I}}%
\end{equation}
is simply a particular matrix multiplication operation on the $\mathcal{Y}%
_{a}^{\mathcal{I}}$. It is well-defined since we are imposing the constraint
that $\mathcal{Y}^{\mathcal{I}}=\mathcal{Y}_{1}^{\mathcal{I}}=\mathcal{Y}%
_{2}^{\mathcal{I}}=...=\mathcal{Y}_{d}^{\mathcal{I}}$. Hence, we can perform a
sum over all the agents in the expression $\mathcal{G}_{\mathcal{IJ}}^{ab}$
$\delta\mathcal{Y}_{a}^{\mathcal{I}}\mathcal{\delta Y}_{b}^{\mathcal{J}}$. The
result is simply the kinetic term for the non-linear sigma model:%
\begin{equation}
\mathcal{G}_{\mathcal{IJ}}^{ab}\text{ }\delta\mathcal{Y}_{a}^{\mathcal{I}%
}\mathcal{\delta Y}_{b}^{\mathcal{J}}\rightarrow\underset{\Sigma
_{\text{agent}}}{\int}d^{d}\sigma\text{ }\sqrt{\det h}\text{ }G_{IJ}%
h^{ab}\frac{\partial y^{I}}{\partial\sigma^{a}}\frac{\partial y^{J}}%
{\partial\sigma^{b}},
\end{equation}
where $G_{IJ}$ measures the infinitesimal proximity between models, and
$h_{ab}$ is the proximity between the agents, as defined by the more general
information metric in equation (\ref{generalized}). Similar considerations
apply for the other contributions to the partition function. This leads us
back to our expression for the partition function in equation (\ref{Zcoll}).
\subsection{Correlations and Inferences}

We can now repurpose many of the methods used in quantum field theory to study
inference schemes of the collective. If we perform the analytic continuation
to a Lorentzian signature agent grid, we can treat the system as a
$(d-1)$-dimensional grid of agents which updates along a grid of time steps.

We specify boundary conditions for the system by introducing a position
dependent chemical potential for the parameters. This is a source in the field
theory:%
\begin{equation}
Z_{\text{coll}}[\mathcal{J}]=\int[\mathcal{D}y]\sqrt{\det G}\exp\left(
-S_{\text{tot}}+\int d^{d}\sigma\text{ }\sqrt{\det h}\text{ }\mathcal{J}%
_{I}(\sigma)y^{I}(\sigma)\right)  .
\end{equation}
Correlation functions of the quantum field theory are given by functional
derivatives with respect to the source:%
\begin{equation}
\left\langle y^{I_{1}}\left(  \sigma_{1}\right)  ...y^{I_{m}}\left(
\sigma_{m}\right)  \right\rangle _{\mathcal{J}} = \frac{1}{Z_{\text{coll}}} \frac{\delta^{m}%
Z_{\text{coll}}[\mathcal{J}]}{\delta\mathcal{J}_{I_{1}}(\sigma_{1}%
)...\delta\mathcal{J}_{I_{m}}(\sigma_{m})}.
\end{equation}
Of particular interest is the behavior of the one-point function $\left\langle
y^{I}\left(  \sigma\right)  \right\rangle _{\mathcal{J}}$ in some asymptotic
limit of the agent parameters $\sigma$. In Euclidean signature this tells us
about the boundary behavior of the collective, and in Lorentzian signature, it
tells us about the late time behavior of the collective.

It is important to stress that our formalism implicitly assumes we are already
close to the true distribution. In other words, we assume that all
fluctuations in the fitting parameters are small.

In some cases the field theory will admit a semi-classical approximation,
corresponding to expansion around a saddle point of $Z_{\text{coll}}$.
Performing a functional derivative of $S_{\text{tot}}$ with respect to
$y^{I}(\sigma)$ yields a classical evolution equation for the $y$'s:%
\begin{equation}
\frac{\delta S_{\text{tot}}}{\delta y^{I}(\sigma)}=\mathcal{J}_{I}(\sigma),
\end{equation}
or,%
\begin{equation}
\nabla^{2}y_{I}-\frac{\partial V}{\partial y^{I}}=\mathcal{J}_{I}(\sigma),
\end{equation}
We view $\mathcal{J}_{I}(\sigma)$ and the $y$ independent part of $\partial
V/\partial y$ as specifying a prior on the guess in the statistical parameter space.

As an illustrative example, consider the case where our guess is a single
centered normal distribution of fixed width. The information metric is the
flat space metric on $\mathbb{R}^{M}$, with entries proportional to the
$M\times M$ identity matrix:%
\begin{equation}
G_{IJ}\propto\text{diag}(1,...,1).
\end{equation}
Let us further suppose that our initial weighting scheme by $w(y)$ specifies a
quadratic potential $V(y)=\frac{1}{2\kappa^{2}}\left(  y-y_{f}\right)  ^{2}$.
This corresponds to an initial weighting scheme by a Gaussian of width
$\kappa$ centered on $y=y_{f}$. All together, this describes a free quantum
field theory with a mass term. The presence of the mass term means that
eventually the configuration will aim to roll towards $y=y_{f}$.

How it gets there is another matter, which is sensitive to the number of
dimensions for the agent space. In $d\geq3$, long range correlations between
agents die off sufficiently quickly that we can use a semi-classical evolution
equation. In $d\leq2$, more care is needed in the nearly massless
limit.

To further illustrate the field theory perspective, let us consider another
example, given by the sum of two normal distributions for outcomes $x\in%
\mathbb{R}
$:%
\begin{equation}
p_{\text{sum}}=\frac{1}{2}\frac{1}{\sqrt{2\pi\alpha^{2}}}e^{-(x-a)^{2}%
/2\alpha^{2}}+\frac{1}{2}\frac{1}{\sqrt{2\pi\alpha^{2}}}e^{-(x-b)^{2}%
/2\alpha^{2}}.
\end{equation}
We hold fixed the width $\alpha$, but vary the parameters $a$ and $b$, so our
statistical manifold is parameterized by the product $\mathbb{R}_{(a)}%
\times\mathbb{R}_{(b)}$. In the limit where $b\simeq a+\varepsilon$, we can
evaluate the leading order form of the information metric:%
\begin{equation}
G_{IJ}=\frac{1}{4\alpha^{2}}\left[
\begin{array}
[c]{cc}%
1 & 1\\
1 & 1
\end{array}
\right]  +\frac{\varepsilon^{2}}{16\alpha^{4}}\left[
\begin{array}
[c]{cc}%
1 & -3\\
-3 & 1
\end{array}
\right]  .
\end{equation}
When $\varepsilon\rightarrow0$, the information metric collapses to a rank one
constant matrix. In the field theory, this means the mode $(a-b)$ does not
propagate, leaving only the center of mass degree of freedom $(a+b)$. The
statistical interpretation is analogous:\ In the limit where the two peaks
coincide, we do equally well to fit the distribution to a single Gaussian; one
of the degrees of freedom has become irrelevant.

\subsection{Stability Under Perturbations}

In this subsection we study the stability of the collective against
perturbations. For simplicity, we shall assume that the potential energy
$V(y)$ only depends on $\sigma$ implicitly through its dependence on $y$. We
shall also assume that the agent space is equipped with a flat metric.

Our aim will be study the growth / dissipation of perturbations to the
original inference scheme as the number of agents $K$ becomes very large.
Alternatively, this can be viewed as asking what happens when the collective
\textquotedblleft changes its mind\textquotedblright\ by perturbing its
initial family of guesses.

We are particularly interested in stable inference schemes. On the one hand,
this means that such perturbations should not drastically change the original
inference. On the other hand, this also means that the collective should be
capable of changing its original inference.

The theoretical tool we use to address the response to perturbations of the
collective is the renormalization group of the field theory. Roughly speaking,
we can partition the original agent space into a set of ever cruder averages
over nearest neighbors. The renormalization group equations track the response
of perturbations as we perform this averaging procedure. Letting
$\lambda_{\text{pert}}$ denote a perturbation of either the background metric
$G_{IJ}$ or the potential $V(y)$, we ask whether $\lambda_{\text{pert}}$ is
amplified or suppressed as the number of agents $K$ grows. In field theory
terms, the scaling behavior defines a beta function:%
\begin{equation}
\beta^{(\lambda)}=\frac{\partial\lambda_{\text{pert}}}{\partial\log K}.
\end{equation}
Stability under perturbations translates to the condition that we arrive at an
interacting conformal fixed point, so in particular we require $\beta
^{(\lambda)} = 0$.\footnote{For earlier work on the potential relevance of scale invariance
in certain inference problems such as applications to perception, see for example
\cite{Bialek:1986it, Bialek:1987qc}, as well as \cite{Nakayama:2010ye}
for a holographic interpretation.}

Rather than delve into a detailed analysis, we will instead present some
heuristic intuition for the different behavior couplings can exhibit by using
--basically classical-- dimensional analysis arguments. The statements we make
receive various quantum corrections, which we shall basically gloss over,
except in subsection \ref{ssec:2D} when we treat the special case $d=2$.

Since we have introduced a length scale $\ell$ which indicates the proximity
between agents on $\Sigma_{\text{agent}}$, scaling with respect to this
parameter will show up when we coarse grain the approximation on the agents.
In particular, this dimensionful scale will also be related to the scaling
behavior of the fitting parameters $y^{I}$. To fix the overall scaling
dimension, we first note that the exponential $\exp(-S_{\text{tot}})$ is only
well-defined provided $S_{\text{tot}}$, the action of the sigma model is
dimensionless. Now, since the integral over the agent space defines a
$d$-dimensional volume Vol$(\Sigma_{\text{agent}})=K\times\ell^{d}$, it
follows that the $y$'s must have appropriate $\ell$ dependence to make
$S_{\text{tot}}$ dimensionless. For example, in the $d=1$ action:%
\begin{equation}
S_{1\text{d}}=\int d\sigma\text{ }\frac{1}{2}\left(  \frac{\partial
y}{\partial\sigma}\right)  ^{2}, \label{1daction}%
\end{equation}
$\left(  \partial y/\partial\sigma\right)  ^{2}$ must scale as $1/\ell$ for
$S_{1\text{d}}$ to be dimensionless. Since each derivative specifies a factor
of $1/\ell$ (c.f. equation \ref{latticederivative}), it follows that $y$
classically scales as $\ell^{1/2}$. More generally, we can consider the case
of a $d$-dimensional agent space. By expanding $G_{IJ}$ around some constant
value, we learn that the classical scaling of $y$ is $y\sim\ell^{\left(
2-d\right)  /2}$.

Now, we are going to use this same dimensional analysis argument to compute
the scaling behavior of perturbations to the original field theory. Though we
do not do so here, one can give more rigorous versions of these arguments.

Consider first the possibility that the collective decides on a different
inference scheme. This corresponds to a perturbation $p_{A}\rightarrow
p_{A}+\delta p_{A}$, which will in turn show up as a change to the original
information metric. In the sigma model, we track this perturbation by
expanding near some point $y_{(0)}^{I}$. Letting $\eta^{I}\equiv y^{I}%
-y_{(0)}^{I}$ denote small variations, we can expand the kinetic term around
this value:\footnote{Actually, if we work with Riemann normal coordinates the
variations begin at quadratic order, with $\delta G_{IJKL}^{(2)}$ proportional
to the Riemann tensor.}
\begin{equation}
\partial y^{I}\partial y^{J}\times G_{IJ}=\partial y^{I}\partial y^{J}%
\times\left(  G_{IJ}^{(0)}+\delta G_{IJK}^{(1)}\eta^{K}+\delta G_{IJKL}%
^{(2)}\eta^{K}\eta^{L}+...\right)  .
\end{equation}
Since $\eta^{I}$ scales in the same way as $y^{I}$, we can work out the
scaling behavior of each of these perturbations as a function of length. Each
subsequent perturbation involves additional factors of $\eta$, so we have
$\delta G^{(m)}\sim\ell^{m\left(  d-2\right)  /2}$. To compute the scaling
behavior of these perturbations with respect to $K$, we recall that the volume
of the agent space Vol$(\Sigma_{\text{agent}})=K\times\ell^{d}$ is being held
fixed throughout the analysis. Hence, we can solve for $\ell$ in terms of $K$,
which implies each such perturbation scales with the number of agents as:
\begin{equation}
\delta G^{(m)}\sim K^{\gamma(m)}\text{ \ \ where \ \ }\gamma(m)=m\times
\frac{2-d}{2d}. \label{gammadef}%
\end{equation}
In other words, in the $K\rightarrow\infty$ limit, these perturbations die off
for $d>2$, while for $d<2$, these terms alter the long distance behavior of
the field theory. For $d=2$, such perturbations could potentially be marginal.

The interpretation in the statistical setting is quite intriguing: If we start
with a given sigma model of statistical inference, we can ask whether the
collective will alter its inference significantly if it makes a change to its
original guessing strategy. These perturbations will show up as variations of
the information metric.

When $d>2$, the collective simply retains its original inference, the inertia
of the group prevents changes.

When $d<2$, on the other hand, we see that each small perturbation has a
dramatic effect on the original inference.

When $d=2$, we see another possibility, namely that such perturbations could
be marginal, and that the collective can smoothly change its inference scheme.
This is a special feature of two-dimensional non-linear sigma models.

For completeness, we perform a similar analysis in the case of a change to the
weighting scheme, i.e. the potential energy $V(y)$. We consider a perturbation
to $w(y)$ of the form:%
\begin{equation}
w(y)\rightarrow w(y)\exp(\delta\lambda_{m}\left\Vert y\right\Vert ^{m}),
\end{equation}
which will show up as a potential term of order $\delta\lambda_{m}\left\Vert
y\right\Vert ^{m}$ in the field theory. Repeating the same scaling analysis,
one learns that $\delta\lambda_{m}$ scales with the number of agents as:%
\begin{equation}
\delta\lambda_{m}\sim K^{1+\gamma(m)}%
\end{equation}
with $\gamma(m)$ as in equation (\ref{gammadef}). Thus, for $d>2$, we see that
a change with $m>2d/(d-2)$ will not alter the original inference. However, for
low enough degree $m$'s, the weighting scheme can significantly alter the
outcome. Similarly, we see that for $d\leq2$, such perturbations are always significant.

\subsection{Two-Dimensional Case \label{ssec:2D}}

Summarizing the previous subsection, we have seen that the case of a
two-dimensional agent space is rather special. In this subsection we discuss
in more detail the conditions for stable inference in two dimensions. From the
perspective of string theory, our discussion of conformal invariance and the
appearance of gravity is not new. How we managed to arrive here
is another story.

We restrict our attention to the well-motivated subcase $V=0$ corresponding to
the Jeffrey's prior. To begin, we also assume a flat agent space metric $h_{ab}$.

Renormalization of 2d non-linear sigma models has received enormous attention
in the string theory literature. See for example \cite{Friedan:1980jm,
AlvarezGaume:1981hn, Callan:1985ia} and \cite{Callan:1989nz} for a review. In
two dimensions there is an additional coupling we can write which also has an
information theoretic interpretation:%
\begin{equation}
S_{\text{2d}}=\frac{N}{2}\underset{\Sigma_{\text{agent}}}{\int}d^{2}%
\sigma\text{ }\sqrt{\det h}\text{ }G_{IJ}\text{ }h^{ab}\frac{\partial y^{I}%
}{\partial\sigma^{a}}\frac{\partial y^{J}}{\partial\sigma^{b}}+\frac{N}%
{2}\underset{\Sigma_{\text{agent}}}{\int}d^{2}\sigma\text{ }iB_{IJ}\text{
}\varepsilon^{ab}\frac{\partial y^{I}}{\partial\sigma^{a}}\frac{\partial
y^{J}}{\partial\sigma^{b}}. \label{S2d}%
\end{equation}
The coupling $B_{IJ}$ is an additional anti-symmetric tensor on the target
space, and $\varepsilon^{ab}$ is a constant anti-symmetric tensor in the
two-dimensional geometry. In the target space, a background value of $B_{IJ}$
specifies a notion of torsion, that is, parallel transport with respect to a
connection $\nabla_{B}$ distinct from the Levi-Cevita connection $\nabla_{LC}$
on $Y_{\text{target}}$. This recovers the use of connections with torsion in
information geometry discussed in section \ref{sec:REVIEW}. Readers familiar
with string theory will note that we have omitted a factor of the dilaton
coupling. This is because the collective works with respect to a fixed and
flat agent space metric.

Assuming the response of the collective is independent of the number of
agents, we arrive at the condition of conformal invariance for the 2d sigma
model. Treating the values of $G_{IJ}$, $B_{IJ}$ as background fields of the
target space geometry, conformal invariance implies the conditions (see e.g.
\cite{Callan:1989nz} for a review):%
\begin{equation}
R_{IJ}-\frac{1}{2}G_{IJ}R=\frac{1}{4}\left(  H_{IJ}^{2}-\frac{1}{6}G_{IJ}%
H^{2}\right)  \text{ \ \ and \ \ }\nabla^{I}H_{IJK}=0 \label{EOM}%
\end{equation}
where $R_{IJ}$ is the Ricci tensor for the information metric, $R$ is the
Ricci scalar, $H_{IJK}=\partial_{I}B_{JK}+\partial_{K}B_{IJ}+\partial
_{J}B_{KI}$. Here we have not written subleading higher derivative
corrections, which are suppressed by powers of $1/N$. Equations (\ref{EOM})
are the Einstein field equations, coupled to a background flux. As is
well-known, they are reproduced by a principle of least action on the
$M$-dimensional target:
\begin{equation}
S_{\text{target}}=\underset{Y_{\text{target}}}{\int}d^{M}y\text{ }\sqrt{\det
G}\text{ }\left(  R-\frac{1}{12}H^{2}\right)  .
\end{equation}
Adding an explicit source of stress energy $T_{IJ}$ to the Einstein equations
corresponds to exposing the collective to a new source of information. In
the semi-classical approximation, we can also identify the graviton as
a fluctuation in the information metric, that is, a fluctuation in
the family of probability distributions adopted by the collective.\footnote{One might speculate on
a connection to entropic gravity \cite{Verlinde:2010hp}, though we do not do
so here.}

From these considerations alone, however, Newton's constant is simply a parameter of the theory.
Indeed, our considerations here are weaker than what would follow from a full string
theory. In particular, the absence of a dilaton equation of motion means that we
cannot fix Newton's constant, and further, that the number of fitting
parameters $M$ is subject to no constraint. The price we pay for this is
that our description will break down at short distances on
$Y_{\text{target}}$, namely \textquotedblleft the string
scale\textquotedblright:%
\begin{equation}
\ell_{\text{min}}\sim1/\sqrt{N}.
\end{equation}
This reflects the underlying $\sqrt{N}$ statistics from the number
of sampled of events, so we typically do not demand more.

If we do demand that our description makes sense below the resolution scale of the
statistics, we need the machinery of string theory. We get a string theory by
summing over all arrangements of the agents, that is, by coupling our sigma model to 2d gravity
on the worldsheet. Then, the dilaton $\Phi$ will automatically appear, and will couple to
the worldsheet curvature via the term $\Phi \cdot \mathcal{R}_{\Sigma}$.
The string frame metric is then the information metric. Stable inference schemes of
the collective then yield the stronger condition of Weyl invariance on a curved
worldsheet. We can eliminate tachyonic modes by passing to the superstring.
Indeed, there is a natural extension of our analysis to the case where the
agent space is a supermanifold.

\section{Information and Compactification \label{sec:COMPACT}}

In this section we switch gears, applying information geometry to study
compactifications of the physical superstring. There are two interlinked questions we
wish to study. First, we want to know whether information
metrics produce accurate approximations to metrics of compactification manifolds.
Second, we want to know how much information an observer would gain by correcting an
incorrect guess of the internal geometry.\footnote{One of the
original motivations for this note was to understand in
information theoretic terms the non-commutative geometry of
F(uzz) theory \cite{FUZZ}, as well as the evidence presented in \cite{TMM,
GMM, TMMII} for the interplay between fuzzy UV cutoffs and an emergent
gravitational sector.}

We shall assume a compactification of the physical superstring of the form
$\mathbb{R}^{3,1}\times Y$, where $Y$ is a Calabi-Yau threefold. We expect
that most of our considerations extend to other K\"ahler threefolds, as can occur in
various F-theory compactifications.

Since $Y$ is Calabi-Yau, and therefore a K\"{a}hler manifold, its metric is
locally controlled by a K\"{a}hler potential $\phi_{Y}$. We can split up the
local coordinates $y^{I}$ into holomorphic and anti-holomorphic coordinates
$y^{i}$ and $\overline{y}^{\overline{j}}$. The metric is then:%
\begin{equation}
ds^{2}=G_{i\overline{j}}^{\text{K\"{a}hler}}\text{ }dy^{i}d\overline
{y}^{\overline{j}}=\partial_{i}\partial_{\overline{j}}\phi_{Y}\text{ }%
dy^{i}d\overline{y}^{\overline{j}}.
\end{equation}
Compare this to the Fisher information metric. Assuming $Y$ is our statistical
manifold of parameters for some probability space $X$, we get distributions
$p(x,\left\{  y\right\}  )$, and an information metric on $Y$:%
\begin{equation}
G_{IJ}^{\text{Fisher}}=-\underset{X}{{\int}}d\mu(x)\text{ }p\frac{\partial
^{2}\log p}{\partial y^{I}\partial y^{J}}=-\left\langle \frac{\partial^{2}\log
p}{\partial y^{I}\partial y^{J}}\right\rangle _{p}.
\end{equation}
Making the further assumption that only the mixed holomorphic and
anti-holomorphic terms contribute, we see that $\log p^{-1}$ defines a
stochastic generalization of the K\"{a}hler potential.

To complete the circle of ideas, we need to find a collection of $p$'s such
that $G_{i\overline{j}}^{\text{Fisher}}$ is a good approximation to
$G_{i\overline{j}}^{\text{K\"{a}hler}}$. Assuming that this can be done, we
can check the inference abilities of an observer by comparing the true
distribution to a nearby guess. This provides a measure of how well an
observer can deduce the local geography\ of a compactification.

\subsection{Construction of Metrics \label{ssec:metric}}

Our strategy for constructing a family of information metrics will be to initially take
$X=\mathbb{C}^{M}$, and an ambient target space $Y=\mathbb{C}^{M}$. Starting
from this, we show how to reduce to an information metric on $Y=\mathbb{CP}%
^{M-1}$, and by embedding a Calabi-Yau in such a projective space, we shall
produce a numerical approximation for a Calabi-Yau metric.

We first construct a family of information metrics on $Y=\mathbb{C}^{M}$ using
the Gaussian:%
\begin{equation}
p(x,\{s\})=\pi^{-M}\det h\text{ }\exp(-h_{\alpha\overline{\beta}}(x^{\alpha
}-s^{\alpha})(\overline{x}^{\overline{\beta}}-\overline{s}^{\overline{\beta}%
})),
\end{equation}
where $x^{\alpha}$ denote holomorphic coordinates for $X=\mathbb{C}^{M}$,
$s^{\alpha}$ denote holomorphic coordinates of the parameter manifold
$Y=\mathbb{C}^{M}$, and $h_{\alpha\overline{\beta}}$ is an $M\times M$
invertible Hermitian matrix with positive eigenvalues. The non-vanishing
entries of the information metric:
\begin{equation}
G_{\alpha\overline{\beta}}^{\text{Fisher}}=\left\langle h_{\alpha
\overline{\alpha}^{\prime}}(\overline{x}^{\overline{\alpha}^{\prime}%
}-\overline{s}^{\overline{\alpha}^{\prime}})(x^{\beta^{\prime}}-s^{\beta
^{\prime}})h_{\beta^{\prime}\overline{\beta}}\right\rangle _{p}=h_{\alpha
\overline{\beta}},
\end{equation}
are generated by the K\"{a}hler potential:%
\begin{equation}
\phi_{(h)}=h_{\alpha\overline{\beta}}s^{\alpha}\overline{s}^{\overline{\beta}%
}.
\end{equation}
Viewing $s^{\alpha}$ as the guess\ of where a probe is located inside the
geometry, we can also adjust the parameters $h_{\alpha\overline{\beta}}$ which
specify the width of the guess. This leads to a much larger parameter space,
with information metric:%
\begin{equation}
ds^{2}=h_{\alpha\overline{\beta}}ds^{\alpha}d\overline{s}^{\overline{\beta}%
}+h^{\overline{\delta}\alpha}h^{\overline{\beta}\gamma}dh_{\alpha
\overline{\beta}}dh_{\gamma\overline{\delta}}.
\end{equation}

To construct information metrics on other parameter spaces, we restrict the
range of parameters of $Y=\mathbb{C}^{M}$. For example, we can define a sphere
$S^{2M-1}$ by restriction on the parameter manifold $Y=\mathbb{C}^{M}$ to the
fixed locus:%
\begin{equation}
h_{\alpha\overline{\beta}}s^{\alpha}\overline{s}^{\overline{\beta}}=r^{2}.
\end{equation}
The restriction to this locus (holding $X=\mathbb{C}^{M}$ fixed) yields a
pullback of the metric on $Y=\mathbb{C}^{M}$ to $S^{2M-1}$. By a similar
token, we can also pull back the probability density on $X=\mathbb{C}^{M}$ to
$S^{2M-1}$. Here, we must be careful to respect the condition that $p$ is a
density, and not a scalar of $X$. The restriction to $\mathbb{CP}^{M-1}$ now
follows by reducing along the $S^{1}$ fiber of the Hopf fibration
$S^{1}\rightarrow S^{2M-1}\rightarrow\mathbb{CP}^{M-1}$, so we produce an
information metric on $\mathbb{CP}^{M-1}$ which agrees with the Fubini-Study
metric generated by the K\"{a}hler potential:%
\begin{equation}
\phi_{(h)}^{FS}=\log\left(  h_{\alpha\overline{\beta}}s^{\alpha}\overline
{s}^{\overline{\beta}}\right)  ,
\end{equation}
with the $s^{\alpha}$ now treated as homogeneous coordinates of $\mathbb{CP}%
^{M-1}$.

A similar construction also applies to the case where $Y$ is Calabi-Yau. The
idea is to fix an ample line bundle $\mathcal{L}$ over $Y$. Since
$\mathcal{L}$ is ample, the space of sections $H^{0}(Y,\mathcal{L}%
)\simeq\mathbb{C}^{M}$ defines an embedding of $Y$ into the projective space
$\mathbb{CP}^{M-1}$:%
\begin{equation}
i:Y\rightarrow\mathbb{CP}^{M-1}\text{ \ \ where \ \ }y\mapsto s^{\alpha}(y).
\end{equation}
This is an embedding to projective space since the $s^{\alpha}$ cannot all
simultaneously vanish. One can view the Calabi-Yau as defined by a set of
polynomial relations amongst the $s^{\alpha}$. This need not be a complete
intersection, since there could be relations amongst the relations (syzygies).
Given the Fubini-Study K\"{a}hler form $\omega_{(h)}^{FS}$ on $\mathbb{CP}%
^{M-1}$, we get a $(1,1)$ form on $Y$ via the pullback $i^{\ast}(\omega
_{(h)}^{FS})$. Hence, each choice of Hermitian
matrix $h_{\alpha\overline{\beta}}$ defines a candidate metric for $Y$.
We view this choice as the inference strategy of a given agent.

To produce a best approximation,\ we follow the numerical procedure for
constructing balanced Calabi-Yau metrics used in \cite{DonaldsonApprox,
Douglas:2006rr}. The optimal choice is fixed under the T-map (see \cite{Tian,
Douglas:2006rr}):%
\begin{equation}
T(h)^{\alpha\overline{\beta}}=\frac{M}{\text{Vol}_{Y}}\int d\text{Vol}%
_{Y}\text{ \ }\frac{\overline{s}^{\overline{\beta}}s^{\alpha}}{h_{\gamma
\overline{\delta}}s^{\gamma}\overline{s}^{\overline{\delta}}}.
\end{equation}
For a Calabi-Yau, the fixed point exists. Now, although this is not yet the
$(1,1)$ form of the Calabi-Yau threefold, repeating the construction for sufficiently
high degree powers of the line bundle $\mathcal{L}^{k}$ defines a
sequence of balanced metrics which eventually converge to the K\"ahler form of
the Calabi-Yau metric (see \cite{Tian, Zelditch}):
\begin{equation}
\frac{1}{k}i_{k}^{\ast}(\omega_{k})\rightarrow\omega_{CY_{3}}%
\end{equation}
as $k\rightarrow\infty$. Observe that in this construction, the Chern class
$c_{1}(\mathcal{L})=[\omega_{CY_{3}}]$, i.e. it picks a ray in the K\"{a}hler
cone which is compatible with geometric quantization of the Calabi-Yau
\cite{Tian}.

The statistical inference interpretation should be clear. We can make a guess
both as to the location $y\in Y$ of the observer, as well as the local profile
of the Calabi-Yau metric, via $h_{\alpha\overline{\beta}}$. A\ sequence of
improved guesses, both in the choice of location, and Hermitian metric
eventually converge on an adequate approximation of the internal geometry.

To illustrate how much an observer can learn by adjusting an incorrect guess
of the local geometry, let us return to the case of distributions on the
ambient space $\mathbb{C}^{M}$. Similar comments apply for all of the pullback
/ restriction maps. Suppose that the true distribution of the observer is
centered at some point $s_{(T)}^{\alpha}$, and that it is designated by a
choice of Hermitian matrix $h_{(T)}$. Then, if our observer makes an incorrect
guess given by some other choice of $s_{(A)}^{\alpha}$ and $h_{(A)}$, the
amount of information it would gain by adjusting its choice is the
Kullback-Leibler divergence:%
\begin{equation}
D_{KL}\left(  p_{T}||p_{A}\right)  =\log\left(  \frac{\det h_{(T)}}{\det
h_{(A)}}\right)  +\text{Tr}\left(  h_{(A)}\cdot h_{(T)}^{-1}-h_{(T)}\cdot
h_{(T)}^{-1}\right)  +\left(  s_{(A)}-s_{(T)}\right)  \cdot h_{(A)}%
\cdot\left(  \overline{s}_{(A)}-\overline{s}_{(T)}\right)  .
\end{equation}
The observer learns both by locating its position, and its width correctly.
The supergravity limit of the compactification corresponds to a sharply
localized observer with $h_{(A)}\rightarrow\infty$, and the small volume
limit corresponds to sending $h_{(A)}\rightarrow0$. In both cases, the
observer learns much by adjusting its width. By the same token, once the
observer get close to the true distribution, it ceases to learn very much
about the internal geometry. In this sense, though there is only one true
distribution, a nearby guess is ``good enough''.

\subsection{Quantum Interpretation}

The information geometry of a Calabi-Yau is also closely connected with tiling
a non-commutative deformation of the geometry with nearly pointlike branes, as
in F(uzz) theory \cite{FUZZ}. In this subsection we show that the Hilbert
space of fuzzy points for $\mathbb{C}^{M}$ induces a family of quantum
information metrics for the Calabi-Yau.

Let us begin by briefly reviewing some background on quantum information
metrics. Given a Hilbert space $\mathcal{H}$ and density matrix $\rho$ which
depends on continuous parameters $y$, we can introduce a quantum analogue of
the Fisher information metric, known as the Helstrom / Bures metric (see
\cite{Helstrom, Bures}):
\begin{equation}
G_{IJ}^{\text{HB}}\equiv\frac{1}{4}Tr_{\mathcal{H}}\left(  \rho\frac{\partial
L_{\rho}}{\partial y^{(I}}\frac{\partial L_{\rho}}{\partial y^{J)}}\right)  ,
\label{infometagain}%
\end{equation}
where we have chosen a convenient normalization convention. Here, $\partial
L_{\rho}/\partial y^{I}$ is implicitly defined via:%
\begin{equation}
\frac{\partial\rho}{\partial y^{I}}=\frac{1}{2}\left(  \rho\frac{\partial
L_{\rho}}{\partial y^{I}}+\frac{\partial L_{\rho}}{\partial y^{I}}\rho\right).
\end{equation}

Extending the cases treated in \cite{FUZZ}, we realize
a non-commutative Calabi-Yau threefold by first constructing
non-commutative $\mathbb{C}^{M}$. Imposing a level constraint and
additional holomorphic constraints leads to non-commutative geometry on
subspaces. Since the restriction maps of the previous section were specified
by holomorphic conditions, we can carry over our construction to the fuzzy case.

In more detail, we first construct the fuzzy Hilbert space of points for
$\mathbb{C}^{M}$. Introduce harmonic oscillators $a^{\alpha}$, $\overline
{a}^{\overline{\beta}}$ which are subject to the commutators:%
\begin{equation}
\left[  a^{\alpha},\overline{a}^{\overline{\beta}}\right]  =h^{\overline
{\beta}\alpha}.
\end{equation}
Next, introduce a vacuum state $\left\vert 0\right\rangle $ such that
$a^{\alpha}\left\vert 0\right\rangle =0$. We then build up a Fock space of
fuzzy points for $\mathbb{C}^{M}$ by acting with creation operators
$\overline{a}^{\overline{\beta}}$ on $\left\vert 0\right\rangle $. Denote this
Hilbert space of states by $\mathcal{H}_{\mathbb{C}^{M}}$. For each
$s^{\alpha}$, introduce a coherent state:%
\begin{align}
\left\vert s\right\rangle  &  =\exp(-h_{\alpha\overline{\beta}}s^{\alpha
}\overline{s}^{\overline{\beta}}/2)\exp(h_{\alpha\overline{\beta}}s^{\alpha
}\overline{a}^{\overline{\beta}})\left\vert 0\right\rangle \\
\left\langle s\right\vert  &  =\left\langle 0\right\vert \exp(h_{\alpha
\overline{\beta}}a^{\alpha}\overline{s}^{\overline{\beta}})\exp(-h_{\alpha
\overline{\beta}}s^{\alpha}\overline{s}^{\overline{\beta}}/2).
\end{align}
These are normalized states, and form an overcomplete basis for $\mathcal{H}%
_{\mathbb{C}^{M}}$. We can introduce a further grading of $\mathcal{H}%
_{\mathbb{C}^{M}}$ by the operator:%
\begin{equation}
\widehat{R}^{2}=h_{\alpha\overline{\beta}}a^{\alpha}\overline{a}%
^{\overline{\beta}}.
\end{equation}
We refer to an eigenspace of $\widehat{R}^{2}$ at eigenvalue $r^{2}$ as the
Hilbert space $\mathcal{H}_{\mathbb{CP}^{M-1}}$ of points for fuzzy
$\mathbb{CP}^{M-1}$ at radius $r$. In what follows we work at large radius.
This is the fuzzy analogue of the K\"{a}hler quotient constraint:%
\begin{equation}
\mathbb{CP}^{M-1}=\left\{  s^{\alpha}\in\mathbb{C}^{M}:h_{\alpha
\overline{\beta}}s^{\alpha}\overline{s}^{\overline{\beta}}=r^{2}\text{ modulo
}U(1)\text{ rephasings}\right\}  .\label{Kquot}%
\end{equation}

The fuzzy Hilbert space of points for a Calabi-Yau $Y$ is obtained by
restricting to states of $\mathcal{H}_{\mathbb{CP}^{M-1}}$ annihilated by the
holomorphic relations $f_{i}(s^{\alpha})=0$ of the commutative geometry:%
\begin{equation}
\mathcal{H}_{Y}=\left\{  \left\vert \psi\right\rangle \in\mathcal{H}%
_{\mathbb{CP}^{M-1}}:f_{i}(a^{\beta})\left\vert \psi\right\rangle =0\right\} ,
\end{equation}
where the $f_{i}$ are themselves possibly subject to further relations (syzygies).

We now produce a quantum information metric for $\mathbb{CP}^{M-1}$, and pull
this back to $Y$ via the embedding $i:Y\rightarrow\mathbb{CP}^{M-1}$. We begin
with a construction of the Fubini-Study metric on $\mathbb{CP}^{M-1}$ by
working with the density matrix for the family of pure states:%
\begin{equation}
\rho(s)=\left\vert s\right\rangle \left\langle s\right\vert .
\end{equation}
Observe that since $\rho^{2}=\rho$, we have $2\delta\rho=\delta L_{\rho}$.

Thus, to compute the quantum information metric it is enough to consider the
variation of $\rho(s)$ as we change $s^{\alpha}\rightarrow s^{\alpha}+\delta
s^{\alpha}$, subject to the variation of the K\"{a}hler quotient constraint in
line (\ref{Kquot}):%
\begin{equation} \label{transversality}
h_{\alpha\overline{\beta}}\delta s^{\alpha}\overline{s}^{\overline{\beta}%
}=h_{\alpha\overline{\beta}}s^{\alpha}\delta\overline{s}^{\overline{\beta}}=0.
\end{equation}
To leading order in $\delta s$, the normalization of $\left\vert
s\right\rangle $ remains fixed because $\delta\exp(-h_{\alpha\overline{\beta}%
}s^{\alpha}\overline{s}^{\overline{\beta}}/2)\simeq O(\delta s^{2})$. Hence,%
\begin{equation}
\delta\rho(s)=\left\vert \delta s\right\rangle \left\langle s\right\vert
+\left\vert s\right\rangle \left\langle \delta s\right\vert +O(\delta s^{2})
\end{equation}
where we have introduced the unnormalized states:%
\begin{equation}
\left\vert \delta s\right\rangle =h_{\alpha\overline{\beta}}\delta s^{\alpha
}\overline{a}^{\overline{\beta}}\left\vert s\right\rangle \text{ \ \ and
\ \ }\left\langle \delta s\right\vert =\left\langle s\right\vert
h_{\alpha\overline{\beta}}a^{\alpha}\delta\overline{s}^{\overline{\beta}}.
\end{equation}
Note that the transversality condition of line (\ref{transversality})
implies $\left\langle s|\delta s\right\rangle =\left\langle \delta s|s\right\rangle =0$.
The quantum information metric therefore reduces to:
\begin{equation}
ds^{2}=\left\langle \delta s|\delta s\right\rangle =h_{\alpha\overline{\beta}%
}ds^{\alpha}d\overline{s}^{\overline{\beta}},
\end{equation}
so we recover the Fubini-Study metric on $\mathbb{CP}^{M-1}$ due to the
K\"{a}hler quotient constraint of line (\ref{Kquot}). Returning to our
discussion in subsection \ref{ssec:metric}, the pullback to $Y$ then gives an
approximation for the Calabi-Yau metric.

\section{Speculative Remarks \label{sec:SPEC}}

In this section we entertain some speculative remarks on other possible
uses of information geometry in the study of superstring theory.
We take an agent to be a D-instanton, that is, the endpoint of an open string.
The D-instanton comes with a local position, and possibly other parameters
such as its width, when it dissolves as flux in another brane. The natural
choice of $Y$ in the supersymmetric setting is simply the moduli space of the brane.

\subsection{Simplistic Holography}

One of the most direct ways to probe the AdS/CFT correspondence is via D-instantons
\cite{Banks:1998nr, Bianchi:1998nk, Dorey:1998qh, Dorey:1999pd}. Indeed, the instanton
density in Yang-Mills theory defines an unnormalized probability distribution, and a corresponding
information metric \cite{HitchinInfo} which provides another way to interpret the AdS/CFT
correspondence \cite{Blau:2001gj}.

Rather than delve into the details of the instanton moduli for a specific gauge theory, we instead
try a more simplistic approach. Given a large $N_{c}$ stack of D3-branes filling $\mathbb{R}^4$ in a consistent
10D geometry, we consider a D-instanton inside the worldvolume of the stack.
We can view the D-instanton as independently sampling the D3-branes
many times with the large number of D$_{-1}$ /
D3 strings. Assuming finite mean and variance for this sampling, the central limit theorem
allows us to approximate the profile of the D-instanton by a Gaussian
on $\mathbb{R}^{4}$. In physical terms, it is actually most
natural to use an unnormalized parametric family of such distributions:
\begin{equation}
p(x,\{y^{(1)},...,y^{(4)},\alpha\})=\frac{L^{2}}{2}\left(
\frac{1}{\pi\alpha^{2}}\right)  ^{4/2}\exp\left[  -\underset{I=1}{\overset
{4}{\sum}}\frac{(x^{(I)}-y^{(I)})^{2}}{\alpha^{2}}\right]  .
\end{equation}
The mean of the distribution tells us
the average position of the D-instanton, and the
width sets a resolution length. The information
metric for this distribution is 5D Euclidean AdS space of radius $L$:%
\begin{equation}
ds^{2}=L^{2}\times\frac{dy_{(I)}dy^{(I)}+d\alpha^{2}}{\alpha^{2}}.
\end{equation}
Similar considerations hold in other dimensions. Observe that our discussion
only required a large number of strings to sample the stack of D3-branes, and
is otherwise insensitive to the details of the gauge theory. It
would be very interesting to see whether a simplistic treatment
along the lines presented here could also reproduce the $S^{5}$
factor of the $\mathcal{N} = 4$ holographic dual.

\subsection{Lorentzian Signature Statistics}

Much of our focus in this work has been on the inference abilities of a
collective of agents probing a Riemannian manifold. For applications to
physics, one should eventually return to Lorentzian signature. This opens up
some new possibilities as well as challenges from the perspective of
information geometry.

To illustrate the main issue, consider again the normal distribution,
but now in Lorentzian signature. We can of course simply write down
a formal Gaussian profile on $\mathbb{R}^{M,1}$:%
\begin{equation}
p_{\text{formal}}(x^{\mu})\equiv(2\pi\alpha_{t}^{2})^{-1/2}\left(  2\pi
\alpha_{t}^{2}\right)  ^{-M/2}\exp\left[  \frac{(t-y_{0})^{2}}{2\alpha_{t}%
^{2}}-\underset{I=1}{\overset{M}{\sum}}\frac{(x^{(I)}-y^{(I)})^{2}}%
{2\alpha_{x}^{2}}\right]
\end{equation}
via the formal analytic continuation from $\alpha_{t}\rightarrow i\alpha_{t}$
(and taking the norm). Since analytic continuation of a D-instanton to Lorentzian signature
defines a tunneling event, we take the unbounded behavior of
$p_{\text{formal}}$ to mean that while the agent eventually settles on a
choice of $y_{i}$ for spatial directions, it continues to move forward in the
temporal direction.

Yet another new feature of Lorentzian signature spacetimes is the possible
existence of null vectors. In information theoretic terms, this means nothing
is learned by moving along the null direction. It would be interesting to
study the consequences for black hole event horizons and cosmological horizons.

A temporal direction in statistical inference also shows up from sequential
updating of the events $e_{1},...,e_{N}$ received by an agent. We can then drop
the distinction between $N$ and $K^{1/d}$ by using a lattice approximation
for a $(d+1)$-dimensional field theory. It would be interesting to connect this to the field
theory interpretation of Bayesian updating in \cite{Bialek:1996kd}.

\section{Conclusions \label{sec:CONC}}

In this note we have studied the interplay between statistical inference and
string theory. We have shown that when a large number of agents form a
collective of nearby guesses, the pooled posterior probability of an accurate
inference by the collective is the partition function of a non-linear sigma
model. Quite surprisingly, stability of the inference scheme against
perturbations requires the collective to arrange along a two-dimensional grid.
Using the well-known fact that the Einstein field equations follow from the
condition of conformal invariance in 2d non-linear sigma models, we found
classical gravity emerge from the condition of stable statistical inference.
We have also taken some preliminary steps in applying methods from information
geometry in the study of the physical superstring.

Developing more practical applications of this formalism would be quite
interesting. For example, one might try to simulate a dynamical updating
strategy for a collective of agents.

Much of our discussion generalizes to the quantum setting, with a quantum
information metric for the non-linear sigma model. It would be interesting to
further study the interpretation in the context of both quantum statistical
inference and string theory.

Finally, we find it rather remarkable that stable computations by a collective
requires a two-dimensional worldsheet, and moreover, that this
yields a theory of gravity. It would be exciting to further develop
an information theoretic formulation of such
\textquotedblleft computables\textquotedblright.

\section*{Acknowledgements}

We especially thank J.J. Heckman Sr., D. Krohn and A. Murugan for helpful
discussions and comments on an earlier draft. We also thank L.B. Anderson,
Y.-T. Chien, S. Detournay and D. Farhi for helpful discussions, and V.
Balasubramanian for encouragement. The work of JJH is supported by NSF grant PHY-1067976.


\bibliographystyle{titleutphys}
\bibliography{InfoSigma}

\end{document}